\documentclass[useAMS,usenatbib]{mn2e}
\usepackage[dvipdfmx]{graphicx}
\usepackage{epsfig}
\usepackage[usenames]{color}

\title[Pulsation modes of RGB OSARG variables]
{On the pulsation modes of OGLE small amplitude red giant variables in
the LMC}
\author[M. Takayama, H. Saio and
Y. Ita]{Masaki Takayama\thanks{E-mail:m.takayama@astr.tohoku.ac.jp},
  Hideyuki Saio 
  and Yoshifusa Ita
\\
Astronomical Institute, Graduate School of Science, Tohoku University, Sendai, Miyagi 980--8578, Japan}
\begin{document}

\date{}

\pagerange{\pageref{firstpage}--\pageref{lastpage}} \pubyear{}

\maketitle

\label{firstpage}

\begin{abstract}
We discuss the properties of pulsations in
the OGLE Small Amplitude Red Giants (OSARGs) in the Large Magellanic Cloud (LMC). 
We consider stars below the red-giant tip in this paper.
They are multi-periodic and form three sequences in the
period-luminosity plane.
Comparing the periods and period ratios with our theoretical models, we have found that
these sequences correspond to radial first to third overtones, and nonradial dipole p$_4$
and quadrupole p$_2$ modes.  
The red-giant branch stars of OSARGs consist of stars have
initial masses of $\sim0.9 - 1.4M_\odot$ which  corresponds to a luminosity range of
$\log L/L_\odot \simeq 2.8 - 3.4$. 
With these parameters, the scaled optimal frequency $\nu_{\rm max}$ for 
solar-like oscillations goes through roughly the middle of the three sequences in the
period-luminosity plane, suggesting  the stochastic excitation is likely the cause
of the pulsations in OSARGs.
\end{abstract}


\section{Introduction}
Recently, the number of known long period variables (LPVs) has increased 
dramatically thanks to the analyses for the extensive photometric data from 
various gravitational microlensing surveys, 
Massive Compact Halo Object (MACHO), 
Optical Gravitational Lensing Experiment (OGLE),  
Exp\'erience de Recherche d'Objets Sombres (EROS), 
and Mirolensing Observations in Astrophysics (MOA) 
\citep[e.g.,][]{woo99,cio01,kis03,sos04,ita04,nod04}.

Analyzing photometric data of red giants brighter than 
$\log L/L_\odot \sim 2.7$ in MACHO survey of the LMC, 
\citet{woo99} found several sequences in the period-luminosity (PL) plane
with periods ranging from about 20 to a thousand days.
They named these sequences A, B, C, E, and D from shorter to longer periods,
in which the sequence C coincides with the known PL relation of Mira variables. 
Later \citet{ita04} found that sequence C' which corresponds to overtone Mira variables 
can be isolated from the sequence B.
 From the light curves, sequence E is attributed to eclipsing binaries \citep{woo99,nic10}.
Sequence D is sometimes called Long Secondary Periods (LSPs) because in
those stars, shorter period pulsations are superposed on a LSP of about 400 -- 1500 days.
Although the origin of the sequence D is discussed by several authors, 
it is not settled yet
\citep[{A historical account on LPVs is given in} e.g.,][]{sos04,sos07}. 
In this paper we do not consider stars on 
sequence D.

\citet{sos04} isolated OGLE small amplitude red giants 
(OSARGs) from the LPVs.
The OSARGs can be divided into two groups; red-giant branch (RGB) stars less luminous
than the giant tip (corresponding to the occurrence of the core helium flash) 
and asymptotic giant branch (AGB) stars.
\citet{sos04} identified three ridges in the PL plane for the RGB OSARGs named b$_1$,
b$_2$ and b$_3$ from longer to shorter periods;
b$_2$ and b$_3$ apparently correspond to sequence B and A of \citet{woo99}, and
B$^{-}$ and A$^{-}$ of \citet{ita04}, respectively.
Furthermore, b$_1$, b$_2$ and b$_3$ correspond to R$_1$, R$_2$, and R$_3$ 
discussed by \citet{kis03} from OGLE-II data. 
Similarly, \citet{sos04} named four AGB OSARGs as a$_1$, a$_2$, a$_3$ and a$_4$. 
In this paper we determine pulsation modes of the RGB OSARGs 
(i.e., b$_1$, b$_2$ and b$_3$ sequences) in the LMC.

\section{Observational properties of RGB OSARGs}
We use periods of RGB OSARGs in the LMC given
in the OGLE-III catalog \citep{sos09}, in which up to three periods are listed for each star.
In the period-luminosity plane, OSARGs are known to form several sequences. 
We adopt the nomenclature of the sequences and the way of selecting RGB OSARGs  
introduced by \citet{sos04}; 
the sequences of  RGB OSARGs are named b$_1$, b$_2$, and b$_3$.

\begin{figure}
\includegraphics[width=0.5\textwidth]{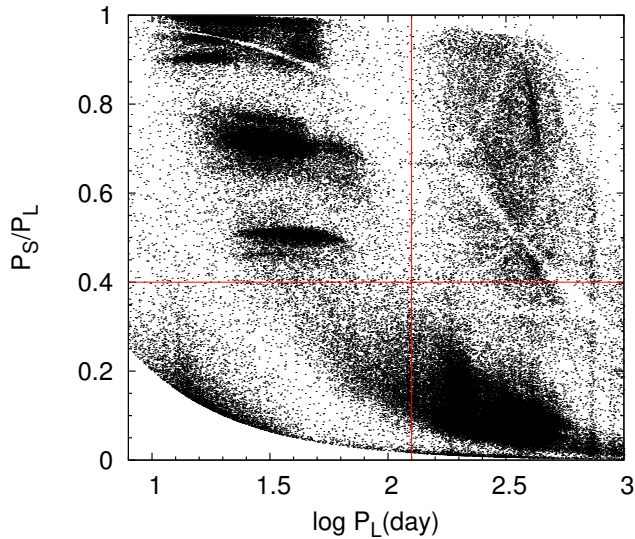}
\caption{All the periods of RGB OSARGs in the catalog of OGLE-III OSARGs 
by \citet{sos09} are plotted in the Period vs period-ratio plane (Petersen diagram).
The ordinate indicates the ratio of the shorter ($P_S$)
to the longer period ($P_L$) for each period pair, 
and the abscissa the logarithmic value of the longer period.
Vertical and horizontal lines are used to exclude stars having periods belong to the 
sequence D (or LSPs).
}
\label{PSD001}
\end{figure}

Figure\,\ref{PSD001} shows a Petersen (period vs period-ratio) diagram 
for all the RGB OSARGs (about 45,500) in the OGLE-III catalog.
In this diagram (and in similar ones shown below), the ordinate indicates 
the ratio of the shorter to the longer period of a period pair, $P_S/P_L$, 
and the abscissa indicates the logarithmic value of the  longer period, $\log P_L$.
Because each star has three periods, each star appears three times in this diagram.
In this figure we see a big dense group in the lower right corner.
The group is formed by the stars associated with 
the long period sequence D; i.e. stars having long secondary periods (LSPs).
The origin of their long periodicity is not clear yet, but
majority of them seem to be irrelevant (or unrelated) to stellar pulsation  
\citep[e.g.,][]{nic09,sos07d,woo04}.

In this paper we do not discuss LSPs. 
To exclude the stars having LSPs from the sample,  we have drawn a horizontal 
line at $P_S/P_L = 0.4$, and a vertical line at $\log P_L=2.1$ in Fig.\,\ref{PSD001}, 
and we have selected stars located only in the upper-left quadrant in this figure.
After excluding stars with LSPs, we have a set of about 8,500 RGB OSARGs in the LMC. 

\begin{figure}
\includegraphics[width=0.5\textwidth]{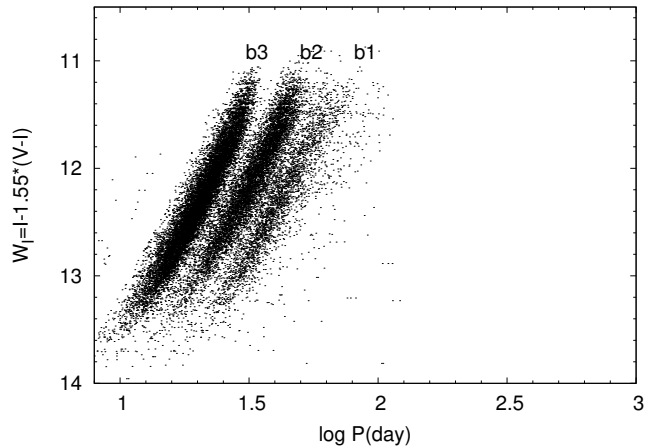}
\caption{Period-$W_{I}$ diagram for our selected sample of RGB OSARGs 
(stars having LSPs are removed),
where $W_I$ is the reddening free Wesenheit index defined by equation (\ref{eq:wi}) 
Three ridges are named b$_1$, b$_2$, and b$_3$ 
(from longer to shorter periods) following the nomenclature introduced by \citet{sos04}}
\label{PLD001}
\end{figure}

The selected stars are plotted in the period-luminosity plane in Fig.\,\ref{PLD001}, where 
the ordinate adopts the reddening free Wesenheit index 
\begin{equation}
W_{I} = I - 1.55(V-I)
\label{eq:wi}
\end{equation}  
with $I$ and $V$ mean magnitudes. 
In this period-luminosity diagram we see three obvious sequences (or ridges). 
They are b$_1$, b$_2$, and b$_3$ sequences from longer to shorter periods,
adopting the nomenclature defined by \citet{sos04}.
We note that the sequence D is clearly removed from our sample.
For the LMC, $W_I \approx 11[mag]$ corresponds to the giant tip at $\log L/L_\odot \approx 3.4$, 
while $W_I \approx 14[mag]$ corresponds to $\log L/L_\odot \approx 2.8$.

\begin{figure}
\includegraphics[width=0.5\textwidth]{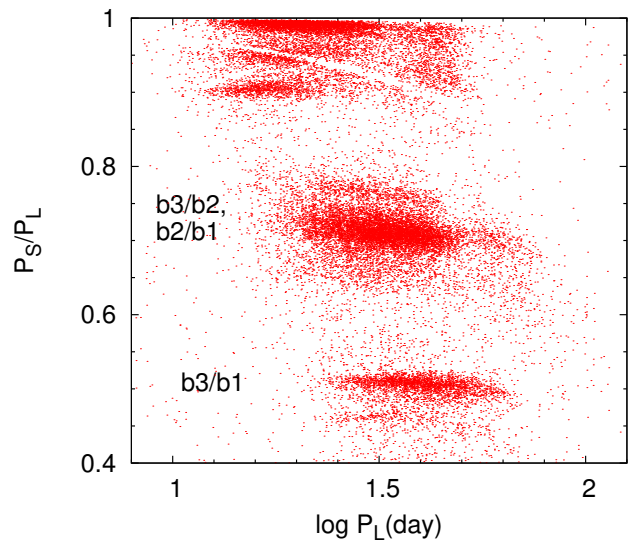}
\caption{Period/Period-ratio (Petersen) diagram for our sample of 
RGB OSARGs (red dots), from which LSPs (sequence D) are removed.
This diagram corresponds to the top-left quadrant of Fig.\,\ref{PSD001}.
Period-ratios of about 0.5 are produced by the ratios of b$_3$/b$_1$,
while ratios of about 0.7 can be attributed to b$_2$/b$_1$ and b$_3$/b$_2$.
The ridges above $\sim0.9$ indicates the presence of nonradial pulsations.
}
\label{PSD002}
\end{figure}

Figure\,\ref{PSD002} shows the Petersen diagram for our selected stars.
This figure is the same as the top-left quadrant of Fig.\,\ref{PSD001}. We note that despite a significant reduction in the number of stars by
excluding LSPs, the property of the distribution in the Petersen diagram is unchanged. 
The structure of the distribution is more clearly seen in Fig.\,\ref{PSD002}. 
We see two major ridges in Fig.\,\ref{PSD002} at $P_S/P_L \approx 0.5$ and $\approx 0.7$;
the former corresponds to the period ratios between  b$_3$ and b$_1$, 
and the latter is formed by the ratios b$_2/$b$_1$ and b$_3/$b$_2$. 
The additional ridges at $P_S/P_L \approx 0.90$ and $\approx 0.95$ are attributed
to the presence of sub-ridges in the P-L plane along b$_2$ and b$_3$ 
as discussed by \citet{sos07}. The less densely populated ridges in Fig.\,\ref{PSD002} at $P_S/P_L \approx 0.45$ and $\approx 0.75$
might be associated with the sub-ridges of b$_2$ and b$_3$. These sub-ridges strongly suggest the presence of nonradial pulsations in OSARGs.

We note that the Petersen diagram (Fig.\,\ref{PSD002}) has an advantage over the period-luminosity
diagram such as Fig.\,\ref{PLD001}  in comparing theoretical results with 
observation to determine pulsation modes. 
Obviously the PL diagram is affected by errors in luminosities and stellar radii,
while only the accuracy in periods for each star matters in the Petersen diagram.  
Therefore, we use Petersen diagrams to determine pulsation modes 
for the period-luminosity sequences, b$_1 -$ b$_3$.

\section{Models}
In order to identify pulsation modes for each of the corresponding
OSARG(RGB) PL relations, we have calculated linear nonadiabatic radial and
non-radial pulsation periods for envelope models along evolutionary tracks.
The evolutionary models were calculated by 
using the \textsc{mesa} stellar evolution code \citep{pax11} for 
initial masses of  0.9, 1.0, 1.1, 1.2, 1.3, 1.4 and 1.6$M_{\odot}$, 
in which wind mass loss is included by adopting Reimers' formula \citep{rei75} 
\begin{equation}
\dot M = - 4\times10^{-13}\eta{(L/L_\odot) (R/R_\odot)\over (M/M_\odot) } 
M_\odot\, {\rm yr}^{-1}
\end{equation}
with $\eta=0.4$.
For our canonical models, we have adopted a chemical composition of
$(X,Z) = (0.71,0.01)$ and a mixing-length of 1.5 pressure scale height.
(We discuss the effects of changing these parameters later in this paper.) 
Opacity tables of OPAL \citep{opal} with low-temperature extension 
by \citet{ale94} were used.

Envelope models for radial and nonradial pulsation analyses were calculated
at selected evolutionary stages of  each evolutionary track.
We set the surface boundary at $\tau$=0.001 for the envelope
  calculations
  with $\tau$ being the optical depth associated with the Rosseland-mean opacity.
  \citet{foxwo82} discussed the uncertainties in the pulsation properties caused by
  various assumptions around the outer boundary of AGB models. 
To check the influence of the particular place of the outer boundary, we have calculated
models imposing the outer boundary condition at  $\tau$=0.01.
We have found that the difference in periods is less than a few percent, 
and that the period ratios are hardly different.
For radial pulsations we set the bottom of an envelope arbitrarily at $r/R\sim1/100$.
For nonradial pulsations we set the inner boundary just below the convective
envelope, suppressing possible coupling with high order core g-modes. 
This is justified because only p-modes completely trapped in the convective envelope
can be excited to observable amplitudes \citep[e.g.,][]{dup09}. 

Linear nonadiabatic analyses for radial and nonradial pulsations were performed 
using the codes described in \citet{sai83} and \citet{sai80}, respectively.
In both cases convection-pulsation coupling is disregarded by 
neglecting the perturbation of the divergence of the convective flux. Convective turnover time in the convective envelope of a typical
model for RGB OSARGs varies from much longer (deep interior) 
to shorter (in the outer layers) than low-order pulsation periods. 
This indicates that we should not trust the stability results of our analyses.  
For this reason we used only periods from these codes disregarding 
the stability results.
We compared our periods and period ratios with
Fig.3 of \citet{xio07}, in which the effect of convection-pulsation 
interaction is included, and found our results to be consistent with their values. 
This confirms that the convection-pulsation coupling hardly affects periods, and
justifies the use of our theoretical periods. 
This is understandable because the nonadiabaticity is not very strong in the RGB models and  the nonadiabatic periods are very close to
the adiabatic ones.
Finally, the fact that the convective turnover time is comparable to the pulsation periods in some layers in the envelope is favorable for the pulsations to be
stochastically excited by turbulent convection.

\section{Comparison with observations}
\subsection{Mode identifications}
The period of a pulsation mode is approximately proportional to 
$1/\sqrt{\overline{\rho}}$ with $\overline{\rho}$ being the mean density \citep[e.g.,][]{cox80},
which depends on the stellar mass and radius. 
Generally the dependence of the periods ratios on these two parameters 
is rather weak. In addition,  Fig.\,\ref{PSD002} indicates 
the period ratios depend only weakly on period.
For that reason,
the period ratios are useful for determining pulsation modes, while pulsation periods are used to determine the appropriate luminosity (or mass) ranges.

\begin{figure}
\includegraphics[width=0.5\textwidth]{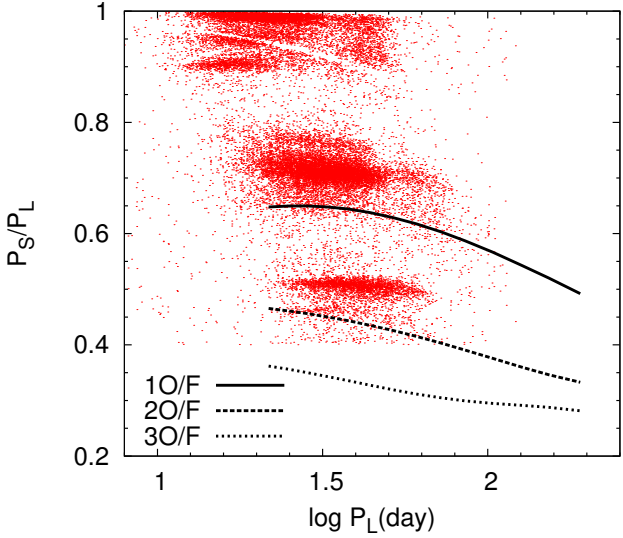}
\caption{Period ratios of 
  the first (1O), second (2O) and third overtone (3O) modes
  to the radial fundamental mode (F) for our 1.1$M_{\odot}$
  models are compared to the observed ratios of OSARGs in the LMC (red dots). 
  The solid line shows the first overtone to fundamental period ratio (1O/F), 
  while the dashed line is for the 2O/F ratio and 
  the dotted line is for the 3O/F ratio. 
  These lines do not fit with any of the ridges in 
  the distribution of OSARG period ratio, indicating that
  fundamental mode are not excited in the RGB OSARGs.}
\label{PSD003}
\end{figure}

\begin{figure}
\includegraphics[width=0.5\textwidth]{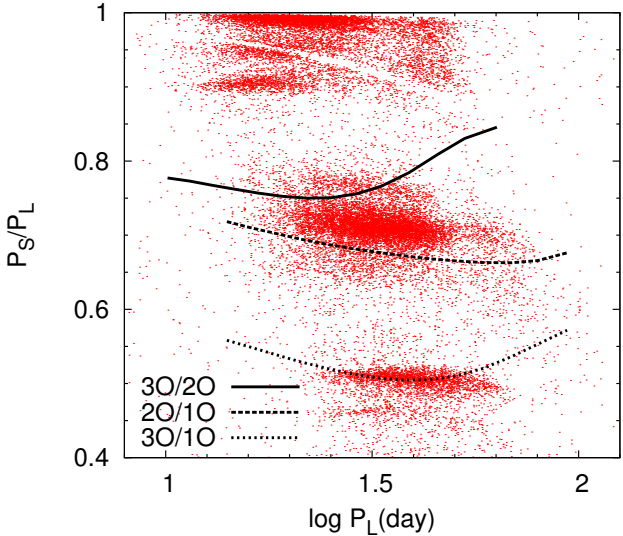}
\caption{The same as \ref{PSD003} but for period ratios without fundamental mode.
The solid, dashed and dotted lines show the ratios between radial overtones as indicated.
The 3O/2O and 2O/1O ratios go through close to the ridge around 0.7,  
while the 3O/1O ratio goes through the ridge at 0.5.
}
\label{PSD004}
\end{figure}

First, we determine which  radial modes are excited in the RGB OSARGs. 
Figs.\,\ref{PSD003} and \ref{PSD004} compare observed period ratios with 
theoretical ones formed by the radial fundamental (F), 1st (1O), 
2nd (2O), and 3rd (3O) overtone periods 
along a part of the evolutionary track ($2.7\le \log L/L_{\odot} \le 3.35$) 
of  an initial mass of $1.1M_\odot$. 
The luminosity range roughly corresponds to that of RGB OSARGs. 
Fig.\,\ref{PSD003} shows the ratios involving fundamental 
mode (denoted as 'F').
Obviously none of the period ratios involving the fundamental mode
agrees with any major ridges around 0.5 or 0.7. 
This indicates that the fundamental mode is not
responsible for any of the period-luminosity sequences, b$_1$, b$_2$, and b$_3$.
For this reason we do not consider the radial fundamental mode any further.
 

Fig.\,\ref{PSD004} shows that the ratio of the third overtone to 
the first overtone (3O/1O) is close to 0.5 in the middle of the luminosity range agreeing with
the b$_3$/b$_1$ ratio.
In addition, the period ratios of 2O/1O  and 3O/2O are 
around 0.7 in the middle of the luminosity range,
although they do not go through the central part of the ridge.
From these comparisons we conclude that the radial first, second, and third overtones
correspond to, respectively,   the b$_1$, b$_2$, and b$_3$ sequences
of the RGB OSARGs.
We note that \citet{dzi10} was unable to explain the period-luminosity sequences b$_2$ and b$_3$ 
by assigning  the first overtone (1O)  to b$_2$ and the second overtone (2O) to b$_3$.
Our identifications differ by one order; i.e., 2O to b$_2$ and 3O to b$_3$.

Fig.\,\ref{PSD004} indicates that the theoretical relations of $1.1M_\odot$ models 
cannot explain the period ratios for the whole period range of the RGB OSARGs.
They agree with the observed ratios 
only in a period range of
$1.5\la \log P(1{\rm O}) \la 1.7$ which corresponds to a
luminosity range of $3.0\la \log L/L_\odot \la 3.15$.
This means that we have to apply models having different initial masses
for different luminosity (hence period) ranges. 
In other words, the running of the PL relations 
must be explained by not the stellar evolution of 
a single mass star, but the evolution of stars with a range of initial
masses. We will discuss different mass models in the next subsection.

\begin{figure}
\includegraphics[width=0.5\textwidth]{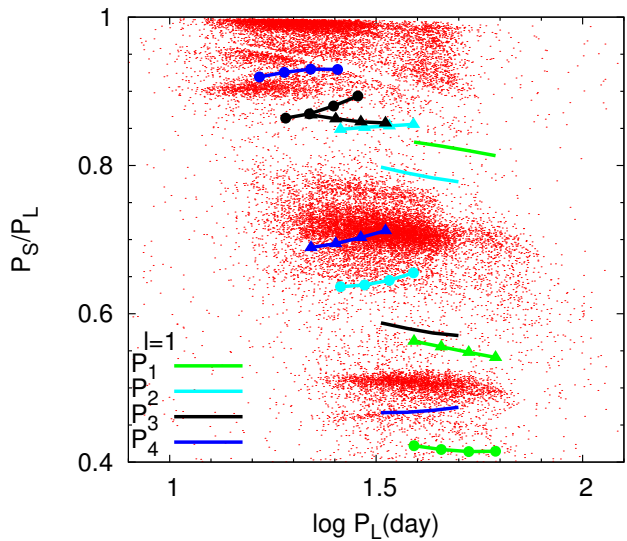}
\caption{The same as \ref{PSD003} but for period ratios of nonradial dipole ($l=1$) 
 p-modes, p$_1 - $p$_4$ to radial modes of 1.1M$_{\odot}$ models.
 Lines (color coded as indicated) without symbols, with triangles, and with circles
 indicate period ratios to radial first (1O), second (2O), and third overtones (3O),
 respectively.
}
\label{PSD006}
\end{figure}

\begin{figure}
\includegraphics[width=0.5\textwidth]{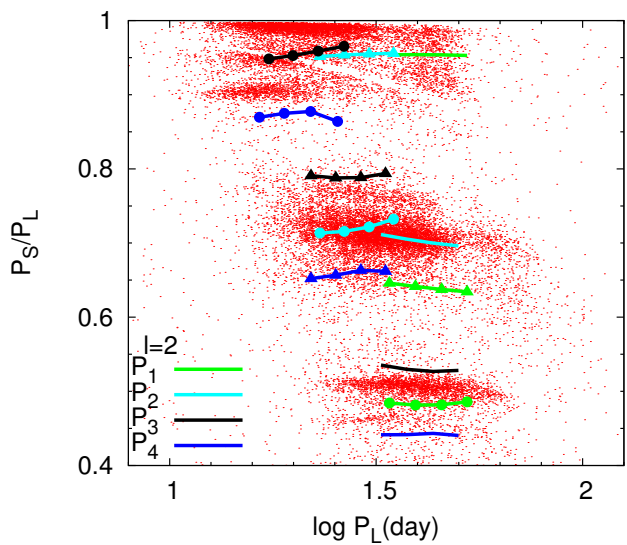}
\caption{The same as Fig. \ref{PSD006} but for nonradial quadrupole 
 ($l=2$) p-modes, p$_1 - $p$_4$.}
\label{PSD007}
\end{figure}

Here, we consider period ratios involving nonradial pulsations in our 
$1.1M\odot$ models in the luminosity range $3.0\la \log L/L_\odot \la 3.15$.
Figs.\,\ref{PSD006} and \ref{PSD007} show period ratios between
dipole ($l=1$) and radial modes,  and quadrupole ($l=2$) and radial modes, respectively,
where solid lines, lines with triangles, lines with filled circles correspond
to the ratios with 1O, 2O, 3O, respectively.
Nonradial modes considered are p$_1 ~\ldots ~{\rm p}_4$ for both $l=1$ and 2. 

Fig.\,\ref{PSD006} indicates that among the dipole modes, only p$_4$ 
is consistent with all the observed ridges in the Petersen diagram.
This mode explains the $P_S/P_L\approx 0.45$ sub-ridge by paring with 1O (b$_1$), 
the $\approx 0.7$ major ridge by paring with 2O (b$_2$), 
and $\approx 0.9-0.95$ by paring with 3O (b$_3$). 
Therefore, we can identity the dipole $(l=1)$ p$_4$-mode as a sub-ridge of b$_3$.

Fig.\,\ref{PSD007} indicates that for the quadrupole ($l=2$) modes, p$_2$-mode 
yields  period ratios 
consistent with the observed ones; ratios with O1 and O3 correspond to the central
ridge at $\approx 0.7$ and the ratio with O2 corresponds to the ridge at $\approx 0.95$.
This means that quadrupole p$_2$-mode corresponds to a sub-ridge of b$_2$.
This figure might also indicate the possible presence of p$_1 (l=2)$ as a sub-ridge
of b$_1$. More data are needed to confirm its presence.   

In summary, we have identified 1O for b$_1$,  2O and p$_2 (l=2)$ for b$_2$, 
and 3O and p$_4 (l=1)$ for b$_3$.
\citet{sos07} argued that each of the sequences b$_2$ and b$_3$ has 
two sub-ridges in addition to a main ridge,
while our mode identifications explain only one sub- and main ridge pair.
Second sub-ridges, if real, might correspond to higher degree ($l\ge3$) modes.
Also, the ratios with the second sub-ridge within each of sequences b$_2$ and b$_3$ 
might be consistent with the period ratios
around 0.98 (e.g., Fig.\,\ref{PSD007}) which we cannot explain with our
mode identifications. 
It is interesting that the main part of the ridge at  $P_S/P_L\approx 0.7$ is reproduced
by the ratios involving nonradial pulsations.
This might indicate that main relations of b$_2$ and b$_3$ sequences 
are formed by nonradial pulsations; p$_2 (l=2)$ and p$_4 (l=1)$, respectively,
rather than radial modes. 

\subsection{Mass ranges}
In the previous subsection we found that radial 1O, 2O, and 3O, and nonradial
p$_4 (l=1)$ and p$_2 (l=2)$ modes are consistent with the
  observed period ratios in the RGB OSARGs, 
but  the evolutionary sequence of a single initial mass
cannot explain the whole range of luminosities
(and hence periods) observed.
We have to consider a mass distribution along the ridges in the PL relations.
\citet{sos07} and \citet{dzi10} used isochrones of several ages for the relations between
mass and luminosity.
Here, we take a more general procedure to obtain the relation between 
mass and luminosity, using the observed PL relations, in particular sequence b$_3$.

In order to compare with theoretical relations we have converted 
the period-magnitude relations shown in Fig.\,\ref{PLD001} 
to period-luminosity relations
in the following way:
$K$ and $J$ magnitudes of each star are obtained from 2MASS photometry data, 
bolometric corrections for the $K$ magnitudes
\begin{equation}
BC_{K}=0.72+2.65(J-K)-0.67(J-K)^{2}
\end{equation}
\citep{bes84} are applied, and
the LMC distance modulus, 18.54 [mag] \citep{tab10} is adopted. 

\begin{figure}
\includegraphics[width=0.5\textwidth]{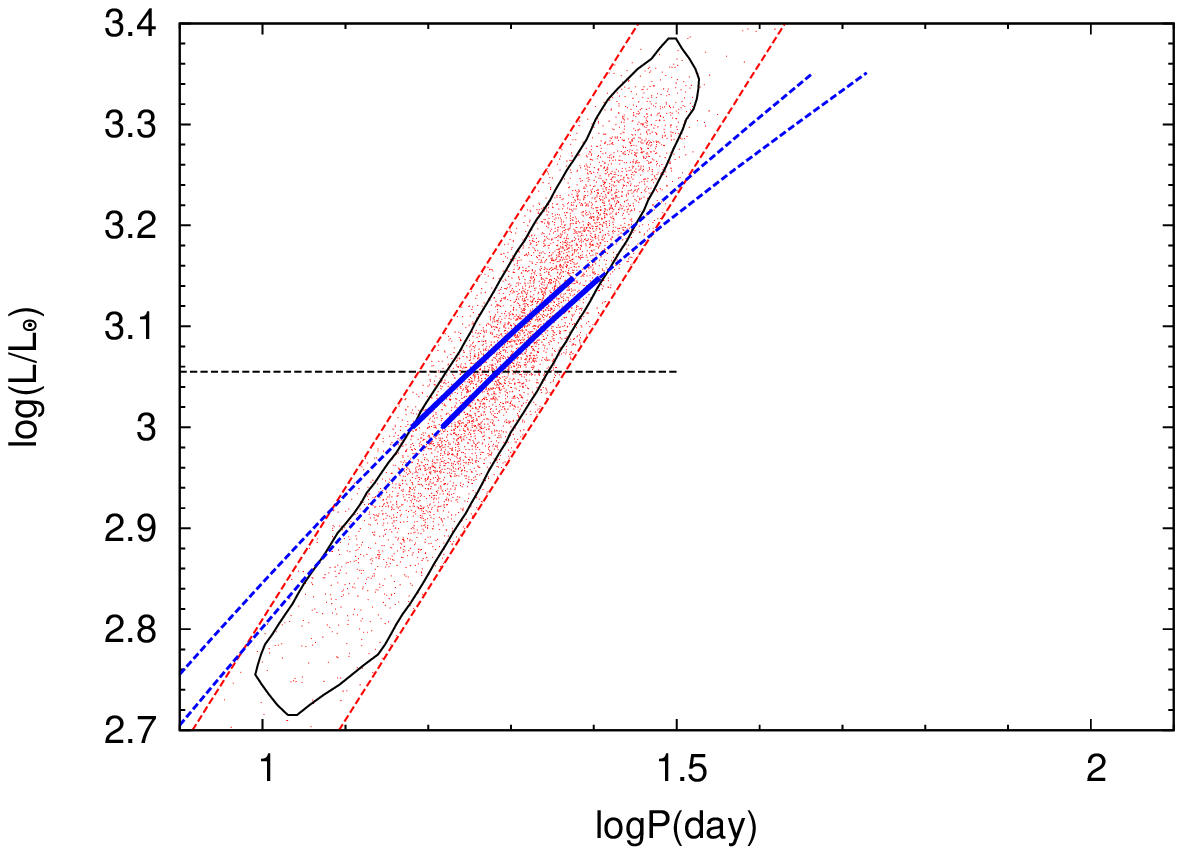}
\includegraphics[width=0.5\textwidth]{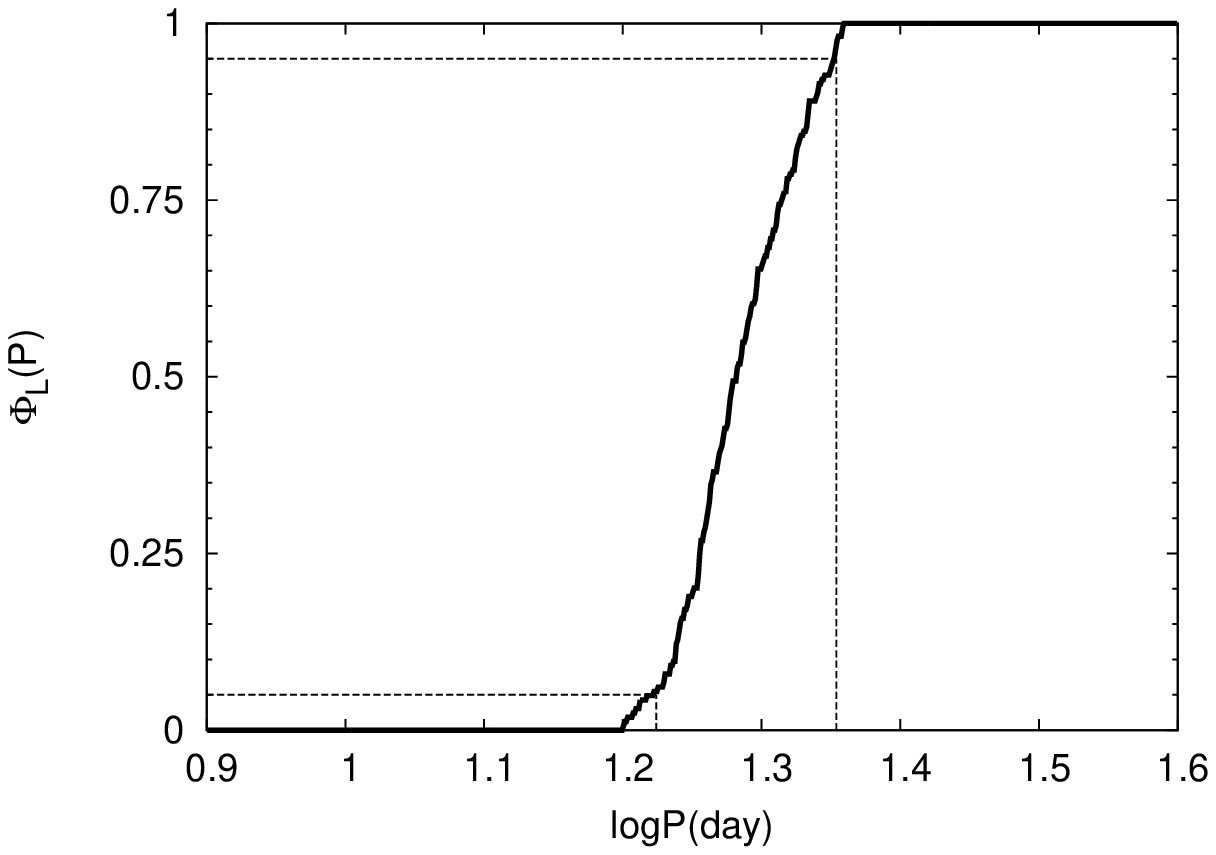}
\caption{The top panel shows period to luminosity relation obtained from
  only stars around sequence b3 (red points; see text for how these stars are
  chosen). The black solid line 
is the quantitative boundary of the sequence b3 defined using
the cumulative distribution function, an example of which is shown in the bottom
panel. 
Dashed (and partially solid) blue lines are relations for the radial 3O and nonradial 
p$_{4}(l=1)$ modes of $1.1M_{\odot}$.
The luminosity range adopted for the mass is shown by the part of solid lines.
The bottom panel shows a cumulative number of stars in a luminosity bin at
$\log L = 3.06$. It is normalized to unity at the longest period is named as
$\Phi_L(P)$.
We have determined boundary of the sequence b3 by assuming that stars lay in the
period range where $0.05 \le \Phi_L(P) \le 0.95$; i.e. the range bounded by 
dotted lines for the luminosity. 
}
\label{cumu}
\end{figure}

For each initial mass we adopt the luminosity range  
such that the theoretical period-luminosity relation 
lies within the `band'  of sequence b$_3$ as shown 
by solid lines in Fig.\,\ref{cumu}.
The boundary of sequence b$_3$ has been determined as follows. 
First we chose stars 
bounded by $1.3 \log P +1.28 \le \log L/L_\odot \le 1.3
\log P+ 1.51$ and $2.7 \le \log L/L_\odot \le 3.4$ 
(dashed red lines in Fig.\,\ref{cumu}) as tentative members of b$_3$.
Then we calculated cumulative numbers as a function of period in each
luminosity bin and normalized it to unity at the longest period considered.
We name this as the normalized cumulative fraction $\Phi_{L}(P)$
(an example is shown in the bottom panel of Fig.\,\ref{cumu}).
We regard the width of the sequence b$_3$ at a given luminosity is bounded 
by periods where $\Phi_{L}(P)=0.05$ and 0.95 (dotted lines in the bottom of 
Fig.\,\ref{cumu}). 
We determine such boundaries for each luminosity bin and taking means with 
adjacent luminosity bins to obtain a smooth curve,
as shown by the black solid line in the top panel of Fig.\,\ref{cumu}. 

Using thus obtained boundary for the sequence b$_3$, we have determined the luminosity
range for each mass.
An example is shown in the top panel of Fig.\,\ref{cumu}.
The two blue lines show
period to luminosity relations for radial 3O and nonradial p$_{4}(l=1)$ modes
of $1.1M_{\odot}$ evolutionary models.
We adopt the appropriate luminosity range for the mass 
such that the both lines are within
the boundary of b$_3$; the part is indicated by solid lines.

\begin{figure}
\includegraphics[width=0.5\textwidth]{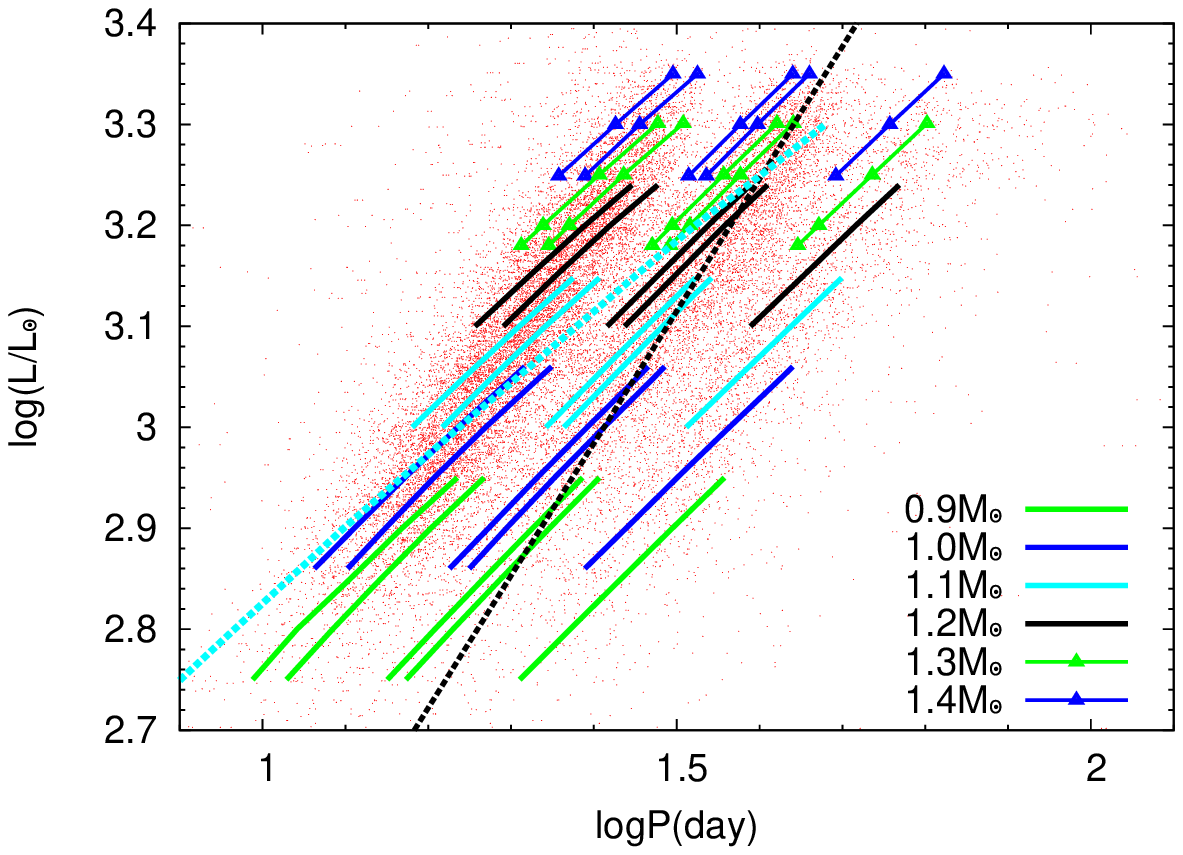}
\includegraphics[width=0.5\textwidth]{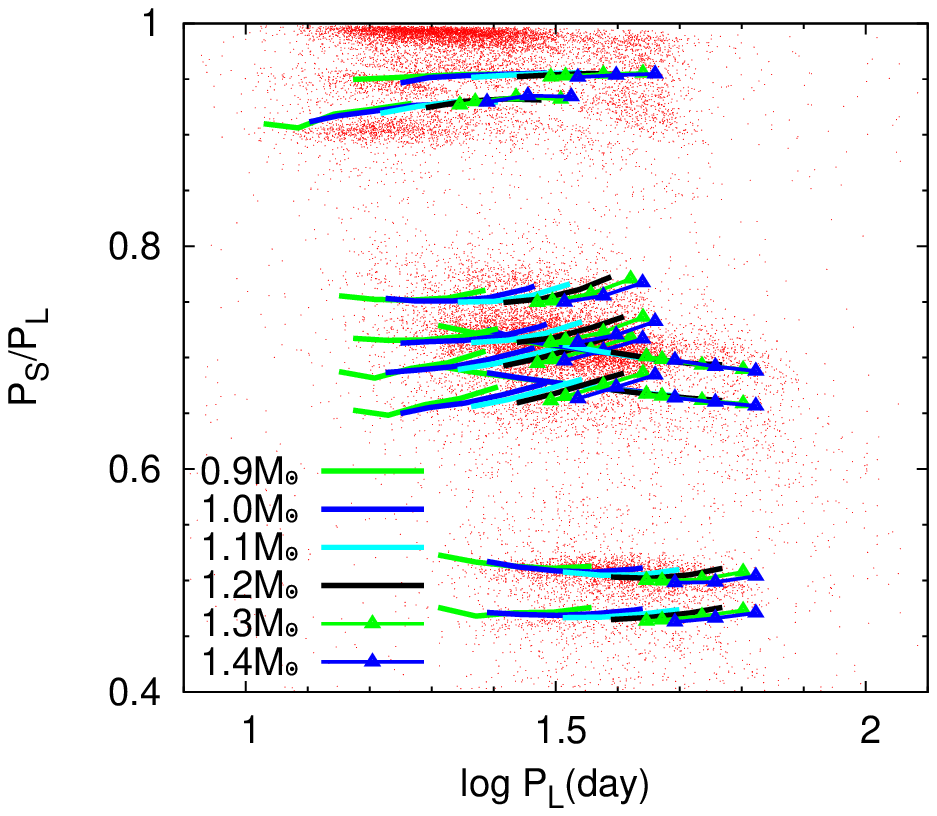}
\caption{Our best models are compared in the period-luminosity plane
  (top panel) and in the period vs period-ratio plane (bottom panel) with RGB OSARGs in the LMC.
Our theoretical relations include radial first to third overtones and the dipole
p$_4 (l=1)$ mode and the quadrupole p$_2 (l=2)$ mode in models 
with masses in the range  $0.9\le M/M_\odot \le 1.4$.
The luminosity range of each mass is determined by fitting with the  
observed b3 sequence as illustrated in Fig.\,\ref{cumu}.
Black dotted line in the top panel shows the scaled relation $1/\nu_{\rm max}$ given 
by equation (\ref{nu_max}) using the mass to mean luminosity relation given
in equation (\ref{eq:ML}) and the corresponding effective temperature.
Cyan dotted line shows $1/\nu_{\rm max}$ computed using the evolutionary track 
of the 1.1$M_{\odot}$ model with $(Z,\alpha)$=(0.004, 1.5).}
\label{fig:best}
\end{figure}

Similarly, we have determined luminosity ranges for models  
with initial masses of  0.9, 1.0, 1.2, 1.3 and 1.4M$_{\odot}$. 
Fig.\,\ref{fig:best} compares thus obtained theoretical period-luminosity
relations (top panel) and period to period-ratio relations (bottom panel) 
with observations.
This figure demonstrates that our models reproduce 
the characteristic distributions of RGB OSARGs in the PL and Petersen
diagrams quite well.
Our models for the RGB
OSARGs consist of the three radial overtones and  the two nonradial
p$_4 (l=1)$, and p$_2 (l=2)$ modes in the evolutionary models having
masses between $0.9$ and $1.4M_\odot$: each mass contributes only 
in the corresponding appropriate luminosity range.
Some other nonradial modes may be involved because clumps in the period-ratios
and ridges in the PL plane are broad.
We note that the need of  different masses for different luminosity ranges comes 
from the fact that  the PL relation of a single mass has a 
inclination slightly different from the observed one. 

The relation between the central value of the luminosity range and mass
can be expressed as an empirical formula
\begin{equation}
\log L/L_{\odot}=0.91(M/M_{\odot})+2.05 
\label{eq:ML}
\end{equation}
obtained by a least square fitting.
We note that the mass-luminosity relation given in this equation
corresponds to ages ranging from about 12.7 Gyr ($0.9M_\odot$)  to 2.6 Gyr
($1.4M_\odot$); 
indicating massive stars are younger than less massive ones.
This means that our mass-luminosity relation is significantly different 
from isochrone relations. 

\citet{dzi10} argued that pulsations in OSARGs should be excited 
stochastically by turbulent
convection because the run of the scaled optimal frequency $\nu_{\rm max}$ 
is consistent with the frequencies of sequences b$_2$ and b$_3$,
where  $\nu_{\rm max}$ is defined as
\begin{equation}
\nu_{max}=\frac{L_{\odot}}{L}\frac{M}{M_{\odot}}\left(\frac{T_{\rm eff}}{T_{{\rm eff}\odot}}
  \right)^{3.5}\times
3050 \mu {\rm Hz}.
\label{nu_max}
\end{equation}
Amplitude is maximum at the frequency $\nu_{\rm max}$ in solar-like oscillations excited 
stochastically by turbulent convection.
The relation was first obtained by \citet{kje95}, scaling the frequency at 
maximum power of the solar oscillations. 
We plot in Fig.\,\ref{fig:best} (top panel; black dotted line) 
the run of $1/\nu_{\rm max}$, 
the optimal period for the stochastic excitation, calculated using equation\,(\ref{eq:ML}) and corresponding effective temperature at each luminosity. 
Also plotted for comparison (cyan dotted line) is the run of $1/\nu_{\rm max}$ 
calculated by using the evolutionary track of the 
 1.1$M_{\odot}$ model with $(Z,\alpha)$=(0.004, 1.5). 
 For both of  the cases, $1/\nu_{\rm max}$ goes
through roughly middle of the PL relations of OSARGs.
This suggests, in agreement with \citet{dzi10}, that pulsations in OSARGs are stochastically excited by 
turbulent convection, and hence kin to the solar-like
oscillations observed in many less luminous red-giants by CoRoT and Kepler
\citep[e.g.,][]{der09, bed10}.
The similarity is also discussed in e.g., \citet{sos07,tab10}. 

\section{Discussion}
\subsection{Effects of metallicity and mixing-length}

\begin{figure}
\includegraphics[width=0.5\textwidth]{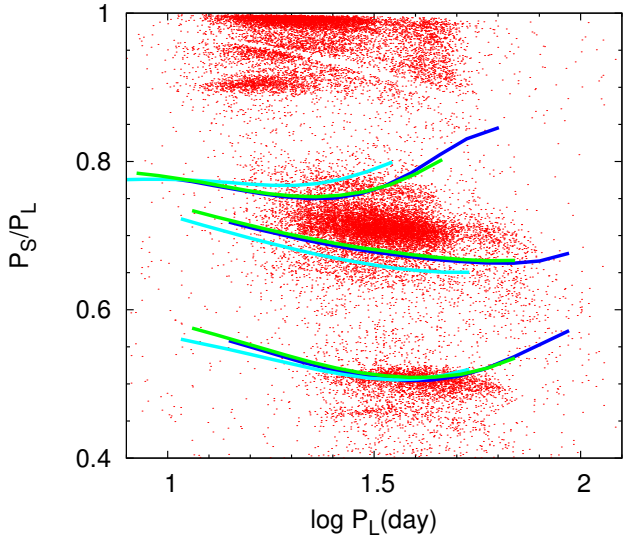}
\caption{The same as Fig.\,\ref{PSD004} but including models with
{\bf $(Z,\alpha)$ } 
different from our standard values.
Blue lines are for the standard models (i.e., the same as in Fig.\,\ref{PSD004}), 
cyan lines are for models with a low metallicity $(Z=0.004)$,
and green lines for models with a longer mixing length $(\alpha=1.8)$. 
}
\label{PSD010}
\end{figure}

In our analyses discussed in the previous sections we have adopted 
a set of parameters $(Z,\alpha) = (0.01, 1.5)$,
where $Z$ is the heavy element abundance and $\alpha$ is 
the ratio of mixing-length to pressure scale height.
We examine here the effect of different choices of $(Z,\alpha)$.
Fig.\,\ref{PSD010} compares the period/period-ratio relations of radial modes 
(3O/1O, 2O/1O, and 3O/2O) for the standard set $(Z,\alpha) =(0.01,1.5)$ with
the $(Z,\alpha) = (0.004,1.5)$ and $(0.01,1.8)$ cases 
for $1.1M_\odot$ models in the luminosity range 
$2.7\le\log L/L_\odot \le 3.35$.
This figure shows that changing metallicity or mixing length hardly changes 
period ratios, although periods themselves shifts considerably.
(Similar results were obtained for different masses in a range of
$0.9\le M/M_\odot\le 1.4$.) 

The period shifts are caused by changes in radius; lower $Z$ or higher $\alpha$
makes the radius smaller (and hence the period shorter) at a given luminosity due to more
efficient energy transport. 
To have the same period we have to adopt a slightly smaller mass compared to
our standard case. 
Although appropriate mass ranges might shift slightly, our mode identifications
for the period-luminosity sequences b$_1$ to b$_3$ are not affected by changing metallicity
or mixing-length.

\subsection{Connection to the solar-like oscillations in G/K giants}
\begin{figure}
\includegraphics[width=0.5\textwidth]{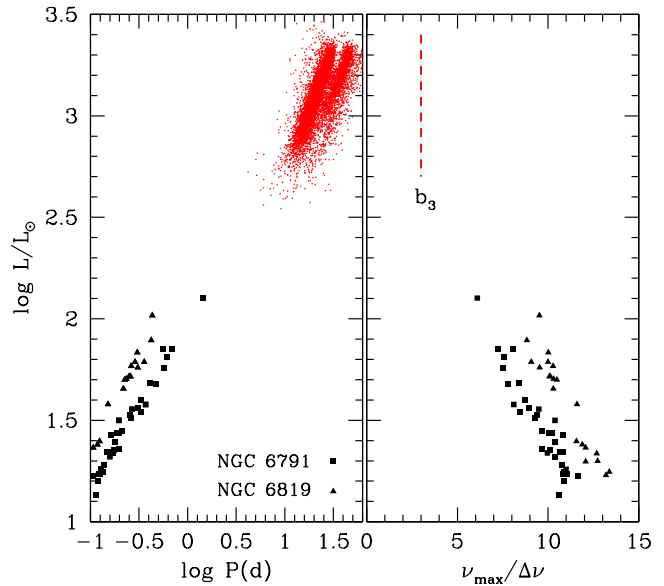}
\caption{Pulsation properties of OSARGs compared with solar-like oscillations of 
open cluster giants in NGC 6791 and NGC 6819 obtained by \citet{bas11} 
from Kepler data.
{\it Left panel} shows luminosity versus period ($1/\nu_{\rm max}$) of maximally excited
mode in the solar-like oscillations of the cluster giants, and
period-luminosity relations of OSARGs in the LMC (red dots), 
where only the primary period for each OSARG is plotted.
{\it Right panel} shows luminosity versus $\nu_{\rm max}/\Delta \nu$  
which gives the approximate number of radial nodes for the maximally excited mode
in the cluster giants. 
The vertical dashed line indicates the number radial nodes of radial 3O mode
which we fit to the sequence b$_3$.  
}
\label{pLnL}
\end{figure}

Another fact supporting the stochastic excitation of OSARG pulsations may be found 
in the period-M$_K$ diagram shown by \citet{tab10} which shows the sequence
of solar-like oscillations in G/K giants extends toward the place where OSARGs are located. 
The left panel of Fig.\,\ref{pLnL} is a similar diagram, 
which shows the relation between luminosity and the period at the maximum amplitude 
($1/\nu_{\rm max}$) for red-giants in the Galactic open clusters 
NGC 6791 and NGC 6819 obtained by \citet{bas11} from  Kepler data.
The difference between NGC 6791 (squares) and NGC 6819 (triangles) 
comes from a difference in the range of stellar masses; stars in the latter cluster systematically more massive
($\sim 1.6 - 2.0 M_\odot$) than those in the former cluster ($\sim 1.0 - 1.5M_\odot$)
according to the seismic mass determinations by \citet{bas11}. 
Also plotted are period-luminosity relations of our sample RGB OSARGs (red dots). 

The right panel of Fig.\,\ref{pLnL} shows $\nu_{\rm max}/\Delta\nu$ of the solar-like
oscillations in the cluster giants, where $\Delta\nu$ means the large separation.
The value of $\nu_{\rm max}/\Delta\nu$ is roughly equal to the number of radial nodes
in the maximally excited mode.
The vertical dashed line indicate the number of radial node (3) for the radial third overtone 
mode which we fit to the sequence b$_3$ of OSARGs.
The radial order of the modes excited in the solar-like oscillations decreases with increasing luminosity,
which is consistent with the fact that only low-order modes are
excited in the OSARGs.

These diagrams indicate that the properties of OSARGs can be
understood as the 
high luminosity (and hence long period) extension of the solar-like oscillations
in G/K giants; in other words OSARGs are likely excited stochastically by turbulent 
convection.
 
\section{Summary}
We studied period-luminosity relations and period ratios of RGB OSARGs in 
the LMC, and found that these relations can be explained by radial first to third overtones
and nonradial p$_4 (l=1)$ and p$_2 (l=2)$ modes in evolving red-giant models
having masses between about $0.9$ and $1.4M_\odot$ for our standard parameters
of $(Z,\alpha) = (0.01, 1.5)$.
Although different choices for the parameters would shift the mass ranges, 
mode identifications are not affected because the period ratios hardly depend on these
parameters.
The sequence b$_1$ is fitted by radial 1O, b$_2$ by radial 2O and nonradial p$_2 (l=2)$,
and b$_3$ by radial 3O and nonradial p$_4 (l=1)$ modes.
The reason why only the two nonradial modes are observed is not clear.

The scaled  $\nu_{\rm max}$ of solar-like oscillations evaluated with our model parameters
goes through roughly the middle of the three ridges populated by RGB OSARGs 
in the period-luminosity plane.
In addition, the period-luminosity relations look like a high-luminosity extension of
the solar-like oscillations recently detected by the Kepler satellite in G/K giants 
of the open clusters NGC 6791 and NGC 6819.
These facts strongly suggest that pulsations in OSARGs are stochastically
excited by turbulent convections.

\section*{Acknowledgments}
This publication makes use of data products from the Two Micron All Sky Survey, which is a joint project of the University of Massachusetts and the Infrared Processing and Analysis Center/California Institute of Technology, funded by the National Aeronautics and Space Administration and the National Science Foundation.

\end{document}